\documentclass[14pt,reprint,twocolumn,longbibliography,
superscriptaddress,
amsmath,
amssymb,
aps,
pra
]{revtex4-2}
\usepackage{graphicx,amsmath,subfigure}%
\usepackage{color,xcolor}
\usepackage[bookmarks=false]{hyperref}
\hypersetup{colorlinks=true,citecolor=cyan,linkcolor=blue,urlcolor=blue,pdfstartview=FitH,bookmarksopen=true}
\usepackage{epstopdf}
\usepackage{float}
\usepackage{ulem}

\begin{document}

\title{One-step implementation of nonadiabatic geometric fSim gate in superconducting circuits}

\author{M.-R. Yun}
\affiliation{School of Physics and Laboratory of Zhongyuan Light, Key Laboratory of Materials Physics of Ministry of Education, Zhengzhou University, Zhengzhou 450052, China}

\author{Zheng Shan}
\email{zzzhengming@163.com}
\affiliation{State Key Laboratory of Mathematical Engineering and Advanced Computing, Zhengzhou 450001, Henan, China}
\author{Li-Li Sun}
\affiliation{College of Physics, Tonghua Normal University, Tonghua 134000, China}
\author{L.-L. Yan}
\affiliation{School of Physics and Laboratory of Zhongyuan Light, Key Laboratory of Materials Physics of Ministry of Education, Zhengzhou University, Zhengzhou 450052, China}
 \affiliation{Institute of Quantum Materials and Physics, Henan Academy of Sciences, Zhengzhou 450046, China}
\author{Yu Jia}
\email{jiayu@zzu.edu.cn}
\affiliation{Key Laboratory for Special Functional Materials of Ministry of Education, and School of Materials and Engineering, Henan University, Kaifeng 475001, China}	
\affiliation{School of Physics and Laboratory of Zhongyuan Light, Key Laboratory of Materials Physics of Ministry of Education, Zhengzhou University, Zhengzhou 450052, China}
\affiliation{Institute of Quantum Materials and Physics, Henan Academy of Sciences, Zhengzhou 450046, China}
\author{S.-L. Su}
\email{slsu@zzu.edu.cn}
\affiliation{School of Physics and Laboratory of Zhongyuan Light, Key Laboratory of Materials Physics of Ministry of Education, Zhengzhou University, Zhengzhou 450052, China}
 \affiliation{Institute of Quantum Materials and Physics, Henan Academy of Sciences, Zhengzhou 450046, China}

\author{G. Chen}
\affiliation{School of Physics and Laboratory of Zhongyuan Light, Key Laboratory of Materials Physics of Ministry of Education, Zhengzhou University, Zhengzhou 450052, China}
\begin{abstract}

Due to its significant application in reducing algorithm depth, fSim gates have attracted a lot of attention. However, during the implementation of quantum gates, fluctuations in control parameters and decoherence caused by the environment may lead to a decrease in the fidelity of the gate. Implementing the fSim gate that is robust to these factors in one step remains an unresolved issue. In this manuscript, we propose a one-step implementation of the nonadiabatic geometric fSim gate composed of a nonadiabatic holonomic controlled phase~(CP) gate and a nonadiabatic noncyclic geometric iSWAP gate with parallel paths in a tunable superconducting circuit. Compared to the composite nonadiabatic geometric fSim gate composed of a nonadiabatic holonomic CP gate and a nonadiabatic geometric iSWAP gate, our scheme only takes half the time and demonstrates robustness to parameter fluctuations, as well as to environmental impacts. Moreover, the scheme does not require complex controls, making it very easy to implement in experiments, and can be achieved in various circuit structures. Our scheme may provide a promising path toward quantum computation and simulation.

\end{abstract}

\maketitle

\section{Introduction}

In the noisy intermediate-scale quantum~(NISQ) era, the implementation of fast and high-fidelity quantum gates has great significance. Although universal gate set can be constructed through arbitrary single-qubit gates and a non-trivial two-qubit gate~\cite{PhysRevA.52.3457}, many algorithms demand a variety of two-qubit gates~\cite{Preskill2018quantumcomputingin}. Replacing an arbitrary two-qubit gate in algorithms requires six to eight single-qubit gates and three CP gates~\cite{KHANEJA200111}. Implementing algorithmic circuits through a series of two-qubit gates directly is of great significance in reducing circuit depth, especially using composite two-qubit quantum gates. Among the most widely used quantum gates, fSim gate has demonstrated its superiority in many NISQ algorithms, including the quantum approximate optimization algorithm~\cite{PhysRevX.10.021067}, linear-depth circuits algorithm for simulating molecular electronic structure~\cite{PhysRevLett.120.110501}, and error mitigation techniques~\cite{Kandala2019}. Therefore, constructing fSim gates has attracted a lot of attention. A combination of an iSWAP gate and a CP gate is used to generate a fSim gate in the standard approach~\cite{PhysRevLett.125.120504}. This leads to a waste of time and decoherence which results in information loss or even collapse and resource waste.  To ensure complete algorithms within the quantum coherence lifetime of quantum systems, one-step construction of high-fidelity fSim gates is highly anticipated.  

Superconducting quantum circuits, due to their scalability, flexibility, and anharmonicity~\cite{doi:10.1146/annurev-conmatphys-031119-050605,Arute2019,Jurcevic_2021,doi:10.1126/science.abg7812}, provide a promising implementation platform for achieving high-fidelity fSim gates. Quantum information in superconducting qubits is lithographically defined. Properties of superconducting qubits such as energy levels, transition frequencies, and anharmonicity are determined by the device parameters in the circuit and can be adjusted according to demand~\cite{PhysRevLett.55.1543}. In addition, states of superconducting qubits can be easily read through non-destructive measurement technology~\cite{RevModPhys.68.1}. Superconducting qubits can be divided into different categories based on the coupling object, and they all have different circuit structures. In recent years, transmon qubit~\cite{PhysRevA.76.042319} is one of the most widespread studied, which can effectively suppress charge noise and is easy to prepare, integrate, and expand~\cite{Fink2011QuantumNI}. However, the second excited state of transmons has a short coherence time. A cross-shape transmon called Xmon~\cite{PhysRevLett.111.080502} which has four legs of the cross is easy to couple maintaining a high level of coherence and has attracted lots of attention~\cite{Barends2014,Kelly2015,Yang2021,Ning2023,Pan2023}.
Recent experiments with superconducting quantum circuits have demonstrated its superiority~\cite{Wang2023,Zhao2021,PhysRevLett.130.030603}. In view of this, the construction of fSim gate using superconducting Xmons is worth investigating.

During the execution of quantum operations, fluctuations of control parameters can seriously affect the fidelity of quantum gates.
Since the geometric phase depends only on the global characteristics of the evolution and is immune to fluctuations of evolution paths, constructing quantum logic gates using geometric phases is seen as one of the effective means to cope with fluctuations of control parameters. In view of the above points, nonadiabatic geometric quantum computation~(NGQC)~\cite{PhysRevLett.89.097902,PhysRevLett.87.097901,PhysRevA.96.052316} and nonadiabatic holonomic quantum computation~(NHQC)~\cite{Sjoqvist_2012,PhysRevLett.109.170501} based on Abelian and non-Abelian geometric phase in two-level and three-level system, respectively, are proposed and have attracted a lot of attentions~\cite{SJOQVIST201665,PhysRevA.94.052310,PhysRevA.92.052302,PhysRevA.94.052310,PhysRevA.98.052315,PhysRevA.108.032601,PhysRevResearch.5.013059,PhysRevA.103.012205,PhysRevApplied.16.044005,PhysRevApplied.16.064040,PhysRevA.102.062410,PhysRevA.106.052610}. However, due to the cyclic evolution conditions of geometric quantum gates, the time of geometric evolution is usually longer than that of dynamical~\cite{Olmschenk_2010,PhysRevLett.114.100503,PhysRevLett.115.043003,doi:10.1126/science.aaf2581}. Recently, to shorten the evolution time, nonadiabatic noncyclic geometric quantum computation (NNGQC) has been proposed and developed~\cite{PhysRevResearch.2.043130,PhysRevA.108.032612,PhysRevA.105.012611,PhysRevA.105.062602}, which goes beyond the limitation of the cyclic condition, makes it more robust against decoherence, and it has been demonstrated in experiment~\cite{PhysRevLett.127.030502}.

Inspired by these, we propose a parallel geometric scheme to implement a fSim gate with the method of NNGQC in the $\{|01\rangle$, $|10\rangle\}$-basis and NHQC in the $\{|11\rangle$, $|02\rangle$,  $|20\rangle\}$- basis, ensuring robustness against control parameter fluctuations. 
 In contrast to the existing theoretical and experimental schemes, the present one has the following characteristics. (i) The gate time required for our scheme with NNGQC+NHQC is one-third of the standard two-step fSim gate and half of the composite nonadiabatic geometric quantum scheme with NGQC+NHQC, which means our scheme is robust against decoherence caused by the environment. (ii) Our scheme is robust to control error and frequency error because of its geometric feature. (iii) The scheme we designed does not require complex controls, making it very easy to implement in experiments, and it can be achieved in various circuit structures. Therefore, our protocol provides a promising strategy for fault-tolerant quantum computation and simulation. 
 
 The manuscript is organized as follows. In Sec.~\ref{Sec2}, we introduce the physical model that used to construct the fSim gate. In Sec.~\ref{Sec3}, we provide a detailed parallel geometric scheme for implementing the fSim gate in one step. In Sec.~\ref{Sec4}, we represent the feasibility and superiority of our scheme through numerical simulation. In Sec.~\ref{Sec5}, we show a circuit that implements the parallel geometric fSim gate with higher adjustability by adding a coupler.
 Finally, we provide an experimental feasibility analysis of our scheme in Sec.~\ref{Sec6} and conclude in Sec.~\ref{Sec7}.

\section{physical model to construct \lowercase{f}S\lowercase{im} gate}\label{Sec2}

\subsection{fSim gate}
The fSim gate, as a combination of iSWAP gate and CP gate, is of great significance in reducing line depth and quantum approximate optimization algorithm. The matrix representation of fSim gate in the $\{|00\rangle,|01\rangle,|10\rangle,|11\rangle\}$ basis given by
\begin{eqnarray}
\rm{fSim}(\vartheta,\Xi)=
\begin{bmatrix}
1&0& 0& 0 \\
0&\cos \vartheta&- i\sin \vartheta & 0  \\
0&-i\sin\vartheta&\cos\vartheta &0   \\
0&0&0 &e^{i\Xi}
\end{bmatrix}.
\end{eqnarray}
 Due to the different frequency requirements of iSWAP gate and CP gate, the fSim gate is generally implemented in two steps, which hinders the implementation of high-fidelity circuits. Moreover, the fluctuation of control parameters can also lead to a decrease in the fidelity of quantum gates. The one-step implementation of robust fSim gates is of great significance in both quantum computation and quantum simulation.

\subsection{Physical Model}

We now proceed to present our scheme based on superconducting circuits. The domain energy of the system is reflected in the $E_J/E_C$ ratio. To reduce the impact of charge noise, which is more difficult to handle than flux noise, and improve the coherence of the system, $E_J\gg E_C$ should be satisfied. In addition, for the convenience of coupling, Xmon qubits are used in our scheme.

\begin{figure}[htbp]
	\centering \includegraphics[width=\linewidth]{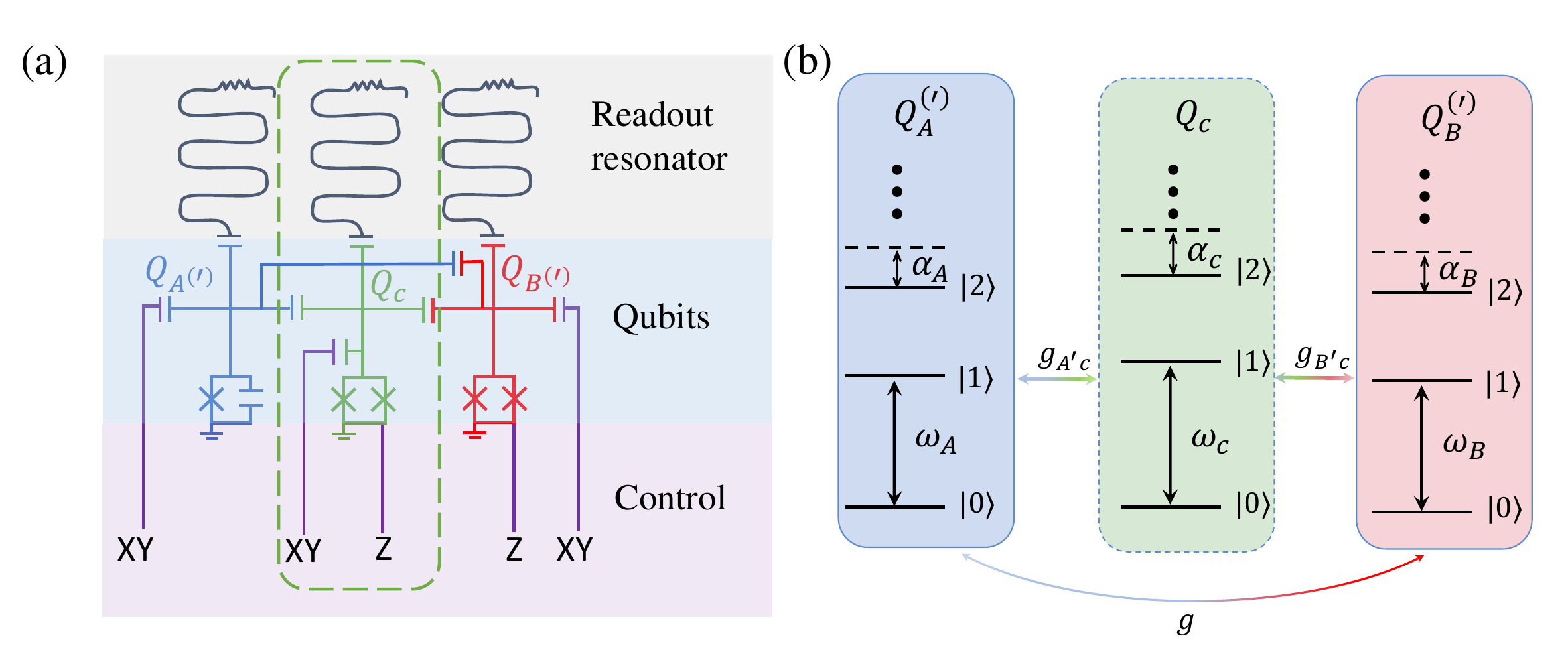}
	\caption{(a) Circuit diagram. For direct coupling, two Xmon qubits $Q_A$ and $Q_B$~(blue and red) are capacitively connected via a capacitor. For indirect coupling, a coupler $Q_c$~(green) is added. $Q_{A^\prime}$ and $Q_{B^\prime}$ are connected by nearest-neighbor coupling with $Q_c$ and next-nearest-neighbor coupling with capacitor. Single-qubit operations can be implemented by local $XY$ control and the frequency of $Q_{B^{(\prime)}}$ and $Q_c$ can be modulated by the magnetic flux with the local $Z$ control line. Each qubit can be measured using a single readout resonator. The capacitance in the superconducting quantum interference device~(SQUID) loop is not shown. (b) Energy structure of each qubit.}
	\label{Fig1}
\end{figure}

We consider two adjacent Xmons $Q_A$ and $Q_B$ are capacitively connected, and the frequency of $Q_B$ can be adjusted by the magnetic flux with the local $Z$ control line, as shown in Fig~\ref{Fig1}(a). The Hamiltonian of this system can be described as
\begin{eqnarray}
\label{eq1}
\hat{H}_{s}&=&	\hat{H}_A+\hat{H}_B+\hat{H}_{\rm int},\notag\\
\hat{H}_A&=& 4E_{CA} \hat{n}_A^2-E_{JA}\cos{\hat{\phi}_A},\notag\\
\hat{H}_B&=& 4E_{CB} \hat{n}_B^2-E_{JB},\notag\\
\hat{H}_{\rm int}&=&4e^2\frac{C_g}{C_1C_2}\hat{n}_1\hat{n}_2,
\end{eqnarray}
where $\hat{H}_A$~($\hat{H}_B$) denotes the Hamiltonian of the individual Xmon $A~(B)$, $\hat{H}_{\rm int}$ is the interaction Hamiltonian of two Xmons, $E_{CA,(CB)}=e^2/2C_{A(B)}$ is the charging energy of the corresponding capacitance, $\hat{n}_{A,(B)}=Q_{A,(B)}/2e$ is the operator of the Cooper-pair number, $E_{JA,(JB)}=I_c \Phi_0/2\pi$ is the energy of the corresponding Josephson with $\Phi_0=h/2e$, $I_c$ is the critical current of the junction, $E_{JB}=E_{JBL}\cos{\hat{\phi}_{BL}}+E_{JBR}\cos{\hat{\phi}_{BR}}$, $E_{JBL(R)}$ is the Josephson energy of the left~(right) junction of $Q_B$. 

The two quantities, $\hat{n}$ and $\hat{\phi}$ obey the canonical commutation relation~\cite{https://doi.org/10.1002/cta.2359}, i.e., $[\hat{\phi},\hat{n}]=i$. With this, in the Xmon regime $E_J\gg E_C$, the Hamiltonian of Eq.~(\ref{eq1}) can be written as ($\hbar=1$)
\begin{eqnarray}
\label{eq2}
\hat{H}_{A}&=& \omega_{A}\hat{a}_{A}^\dagger \hat{a}_{A}-\frac{\alpha_{A}}{2}\hat{a}^\dagger_{A}\hat{a}^\dagger_{A}\hat{a}_{A}\hat{a}_{A},\notag\\
\hat{H}_{B}&=& \omega_{B}\hat{a}_{B}^\dagger \hat{a}_{B}-\frac{\alpha_{B}}{2}\hat{a}^\dagger_{B}\hat{a}^\dagger_{B}\hat{a}_{B}\hat{a}_{B},\\
\hat{H}_{\rm int}&=& g(\hat{a}^\dagger_A\hat{a}_B+\hat{a}_A\hat{a}^\dagger_B-\hat{a}^\dagger_A\hat{a}^\dagger_B-\hat{a}_A\hat{a}_B),\notag
\end{eqnarray}
where

 \begin{eqnarray}
 \label{Ec}
 \hat{a}^\dagger_{A(B)}&=&\frac{1}{\sqrt{2\omega_{A(B)}}}(\sqrt{8E_{CA(B)}} i\hat{n}_{A(B)} +\frac{\omega_{A(B)}\hat{\phi}_{A(B)}}{\sqrt{8E_{CA(B)}}})  ,\notag\\
  \hat{a}_{A(B)}&=&\frac{1}{\sqrt{2\omega_{A(B)}}}(-\sqrt{8E_{CA(B)}} i\hat{n}_{A(B)} +\frac{\omega_{A(B)}\hat{\phi}_{A(B)}}{\sqrt{8E_{CA(B)}}})  ,\notag\\
\omega_{A(B)}&=&\sqrt{8E_{JA(B)}E_{CA(B)}}-E_{CA(B)},\notag\\
\alpha_{A(B)}&=&E_{CA(B)},\notag\\
g&=&\frac{1}{2}\frac{C_{AB}}{\sqrt{C_AC_B}}\sqrt{\omega_A\omega_B}.
\end{eqnarray}

Energy structure of each qubit is shown in Fig.~\ref{Fig1}(b).
The Hamiltonian under the rotating-wave approximation~(RWA) including levels with two excitations for the system in Fig.~\ref{Fig1} can be written as
\begin{eqnarray}
\label{H1}
\hat{H}_1=
\begin{bmatrix}
\omega_{00}&0& 0& 0&0 &0   \\
0&\omega_{01}&0 & g&0 &0   \\
0&0&\omega_{02} &0 &\sqrt{2}g & 0   \\
0&g&0 &\omega_{10} &0 &0   \\
0&0&\sqrt{2}g &0 &\omega_{11} &  \sqrt{2}g  \\
0&0& 0& 0&\sqrt{2}g&\omega_{20} 
\end{bmatrix}
\end{eqnarray}
in the $\{ |00\rangle,~|01\rangle,~|02\rangle,~|10\rangle,~|11\rangle,~|20\rangle \}$-basis, where the bare state of Xmons  A and B is denoted as $|n_1n_2\rangle =|n_1\rangle  \otimes|n_2\rangle $ ($n_i\in 0, 1,2$). We can observe that there is coupling between $|01\rangle $and $|10\rangle $, as well as between $|11\rangle \ |02\rangle $ and $|20\rangle $ simultaneously, so it is possible to achieve both iSWAP and CP gates simultaneously.

\begin{figure}[htbp]
	\centering \includegraphics[width=\linewidth]{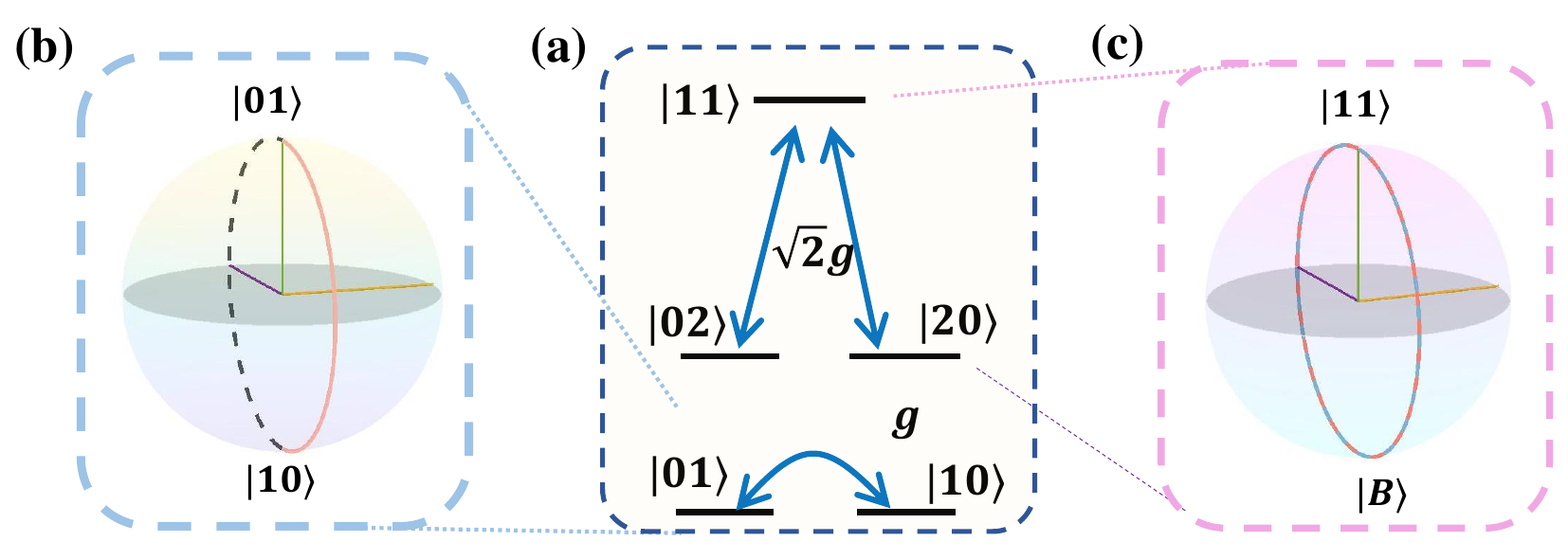}
	\caption{ (a) Energy level diagrams of the double-excitation manifold and the single-excitation manifold in the case of direct coupling when performing the fSim gate. Bloch-sphere representation of the relevant energy in the single-excitation manifold (b) and the double-excitation manifold (c).}
	\label{Fig2}
\end{figure}

\section{parallel implementation of \lowercase{f}S\lowercase{im} gate with NNGQC+NHQC }\label{Sec3}
The fSim gate is a combination of an iSWAP gate and a CP gate. Generally speaking, implementing iSWAP and CP gates in superconducting circuits requires different conditions. When implementing the iSWAP gate, it is usually necessary to have two frequencies of two logical qubits equal to each other, to achieve resonance between $|01\rangle$ and $|10\rangle $. However, when implementing the CP gate, the frequency difference between $|01\rangle$ and $|10\rangle$ needs to be $\alpha$, i.e., the level anharmonicity of each qubit. Therefore, the fSim gates are generally completed in two steps.
In this section, we will introduce how to simultaneously implement these two two-qubit gates in one step.

To implement the fSim gate in one step, three energy levels of each transmon are considered, and the parameters of transmons can be set as $\omega_{A}=\omega_{B}$, $\alpha_{A}=\alpha_{B}$, the energy level structure and coupling are shown in Fig.~\ref{Fig2}(a).
Considering the preparation process, the level anharmonicity of each qubit is only related to the energy of the capacitor, making it easy to achieve equality, but for the frequency, it is very difficult to make the frequencies of two qubits completely equal, Xmon B is designed to be frequency adjustable. The $ |0 \rangle_B \rightarrow |1\rangle_B$ transition frequency of the $Q_B$ as a function of flux bias is shown in Appendix.~\ref{a1}. 
\subsection{NHQC implementation of the CP part}
First, we will demonstrate how to obtain a nonadiabatic holonomic phase on $|11\rangle$~\cite{PhysRevA.92.052302,SJOQVIST201665}.
When $\omega_{20}=\omega_{02}$, the interaction between $|11\rangle,\ |02\rangle $ and $|20\rangle$ can be regarded as a three-level structure with the same detuning. Under these parameter conditions, the Hamiltonian  Eq.~(\ref{H1}) after applying the rotating frame with transform operator $V(t)=\exp[{-i(\omega_{11}|11\rangle\langle11|+\omega_{02}|02\rangle\langle 02|+\omega_{20}|20\rangle \langle 20 |)t}]$, and ignoring rapidly oscillating terms in the RWA can be written as
\begin{eqnarray}
\label{eq5}
	\hat{H}_2=\Delta|11\rangle\langle 11| +2g(|11\rangle\langle B|+\rm H.c.),
\end{eqnarray}
with $\Delta=\alpha_{A}=\alpha_{B}$, $|B\rangle=\frac{1}{\sqrt{2}} (|02\rangle+|20\rangle)$. 
There is a dark state $|D\rangle=\frac{1}{\sqrt{2}}(|02\rangle-|20\rangle)$ of Eq.~(\ref{eq5}), which is decoupled from the system completely. We set 
\begin{eqnarray}
    \Delta&=&2\Omega\sin \gamma,\notag\\
    g&=&\frac{1}{2}\Omega\cos\gamma,
\end{eqnarray}
Then, 
\begin{eqnarray}
\hat{H}_2=&&\Omega\sin\gamma(|11\rangle\langle11|+|B\rangle\langle B|)+\Omega[\cos \gamma|B\rangle\langle11|+\rm H.c.]\notag\\
&&+\Omega\sin\gamma(|11\rangle\langle11|-|B\rangle\langle B|).
\end{eqnarray}
There are only states $|11\rangle$ and $|B\rangle$ are coupled, in the basis of $|11\rangle$ and $|B\rangle$, we can map $|11\rangle\langle 11|+|B\rangle\langle B|\rightarrow \mathbb{I},\ |11\rangle\langle B|+|B\rangle\langle11|\rightarrow\sigma_x,\ |11\rangle\langle 11|\rightarrow -|B\rangle\langle B|\rightarrow\sigma_z$ with $\mathbb{I}$ is the identity matrix, $\sigma_x$ and $\sigma_z$ are the Pauli matrices. Based on this, 
\begin{eqnarray}
    \hat{H}_2=\Omega\sin\gamma \mathbb{I}+\Omega(\cos\gamma\sigma_x+\sin\gamma\sigma_z).
\end{eqnarray} 
When the evolution time $\tau=\pi/\Omega$, the evolution operator of the three-level system can be represented as
\begin{eqnarray}
	U_2(\tau,~0)=e^{-i\phi}|B\rangle\langle B|+e^{i\phi}|11\rangle\langle11|+|D\rangle\langle D|,
 \label{evolution}
\end{eqnarray}
where $\phi=\pi \sin \gamma$.
After undergoing a cycle of evolution with $\tau_2=2\pi/(16g^2+\Delta^2)$, $|11\rangle$ in the computational subspace will obtain a geometric phase. 

Then, we will examine whether this evolution is holonomy, the holonomy transformation should satisfy two conditions, (i) the evolution of the subspace is cyclic, and (ii) there is no dynamical phase in this cyclic evolution. For condition~(i), in the subspace spanned by $\{|B\rangle,\ |11\rangle \}$, from the evolution operator in Eq.~(\ref{evolution}), we can see condition~(i) is satisfied. For condition~(ii), $\langle m|\hat{H}_2|r\rangle$, where $m,r\in\{11,B\}$, it easily to verify. That is to say, the gate in subspace $\{|11\rangle,|02\rangle,|20\rangle\}$ is a holonomy gate.

\subsection{NNGQC implementation of the iSWAP part}
At the same time, under these parameters set, $|01\rangle $ and $|10\rangle$ can achieve resonance interaction, the energy level structure and coupling are shown in Fig.~\ref{Fig2}(a). 
To obtain tunable coupling between $|01\rangle\leftrightarrow|10\rangle$, the flux bias can be set as magnetic flux~\cite{PhysRevApplied.13.064012}, and the frequency of Xmon B in Appendix~\ref{a1} can be re-written in the form of 
\begin{eqnarray}
    \omega_B(t)=\omega_B+\epsilon\sin (\nu t+\varphi),
\end{eqnarray}
where $\nu$ and $\varphi$ indicate the frequency and phase of the modulated field, respectively. 
Then, using the Jacobi-Anger identity 
\begin{eqnarray}
    \exp[i\beta \cos(\nu  t+\varphi)]=\sum_{m=-\infty}^{+\infty} i^m J_m(\beta)\exp[i m (\nu t +\varphi)]\notag\\
\end{eqnarray}
with $J_m(\beta)$ being Bessel functions of the first kind with $\beta=\epsilon/\nu$. 

When $\Delta=\omega_A-\omega_B=\nu$, the Hamiltonian in the basis $|01\rangle$ and $|10\rangle$ with RWA in the interaction picture can be written as
\begin{eqnarray}
    \hat{H}_r=\left(\begin{array}{cc}
         0 &  \mathcal{G} \ e^{-i(\varphi-\pi/2)}\\
        \mathcal{G}\  e^{i(\varphi-\pi/2)} & 0
    \end{array}\right),
\end{eqnarray}
where $\mathcal{G}=J_1(\beta)g$.

To shorten the evolution time, we will now use the method of NNGQC to implement the iSWAP part. We choose a set of auxiliary states 
\begin{eqnarray}
        |\phi_1(t)\rangle&=&\cos\frac{\alpha(t)}{2}e^{-i\frac{\eta}{2}}|01\rangle+\sin\frac{\alpha(t)}{2}e^{i\frac{\eta}{2}}|10\rangle\notag\\
     |\phi_2(t)\rangle&=&\sin\frac{\alpha(t)}{2}e^{i\frac{\eta}{2}}|01\rangle-\cos\frac{\alpha(t)}{2}e^{i\frac{\eta}{2}}|10\rangle.
\end{eqnarray}
 Then, we will take them into the von Neumann equation~\cite{PhysRevLett.123.100501}
 \begin{eqnarray}
     \dot{\Pi}_m(t)=-i[H(t),\Pi_m(t)],
 \end{eqnarray}
where $\Pi_m(t)=|\phi_m(t)\rangle\langle\phi_m(t)|$.
Then, we can get restriction equations for two Hamiltonian parameters
\begin{eqnarray}
    \mathcal{G}&=&\frac{\dot{\alpha}}{2\sin[\varphi(t)-\pi/2-\eta(t)]}\notag\\
    \varphi(t)&=&\eta(t)-\arctan\left[\frac{\dot{\alpha}(t)\cot(\chi(t)}{\dot{\eta}(t)}\right]+\frac{\pi}{2}.
\end{eqnarray}
The evolution operator is
\begin{widetext}
    \begin{eqnarray}
U_r(T,0)&=&e^{i\gamma}|\phi_1(\tau)\rangle\langle\phi_1(0)|+e^{-i\gamma}|\phi_2(\tau)\rangle\langle\phi_2(0)|,\notag\\
 &=&\left(\begin{array}{cc}
e^{i\frac{\eta_-}{2}}(\cos\gamma\cos \frac{\alpha_-}{2}+i \cos \frac{\alpha_+}{2}\sin\gamma)&   e^{-i\frac{\eta_+}{2}}(-\cos\gamma\sin \frac{\alpha_-}{2}+i \sin \frac{\alpha_+}{2}\sin\gamma)\\
 e^{i\frac{\eta_+}{2}}(\cos\gamma\sin \frac{\alpha_-}{2}+i \sin \frac{\alpha_+}{2}\sin\gamma)&  e^{i\frac{\eta_-}{2}}(\cos\gamma\cos \frac{\alpha_-}{2}-i \cos \frac{\alpha_+}{2}\sin\gamma)
    \end{array}\right),\notag\\
\end{eqnarray}
\end{widetext}
where $\alpha_\pm=\alpha(T)-\alpha(0)$, $\eta_\pm=\eta(T)=\eta(0)$, and $\gamma=\gamma_d+\gamma_g$ is the total phase including a geometric phase $\gamma_g=i\int_0^\tau\langle\phi_1|\dot{\phi}_1\rangle\rm dt=\int_0^\tau\frac{1}{2}\dot{\eta}\cos\chi \rm dt$ and a dynamical one $\gamma_d=-\int_0^\tau\langle\phi_1|\hat{H}_r |\phi_1\rangle \rm  dt=-\int_0^\tau \mathcal{G}\cos\alpha(t) \cos[\varphi(t)-\pi/2-\eta(t)]\rm dt$. To obtain a pure geometric phase, we set $\varphi(t) -\eta(t)=0$. Therefore, the total phase $\gamma=\gamma_g=\int_{\eta(0)}^{\eta(\tau)}\int_{\alpha(0)}^{\alpha(\tau)}\frac{1}{2}\sin\alpha\rm d\alpha d\eta$ is half of the solid angle enclosed by the trajectory and the geodesic connecting the initial and final points~\cite{PhysRevResearch.2.043130}. When $\eta_+=-\pi$, $\eta_-=0$, $\alpha_+=0$, $\alpha_-=-\pi$, and $\gamma=\pi$, iSWAP operation can be obtained, and the evolution time of such a nonadiabatic noncyclic process is $\tau_1=\pi/2\mathcal{G}$. 

\subsection{Implementation of the fSim gate with parallel paths}
\begin{figure}[htbp]
 	\centering \includegraphics[width=\linewidth]{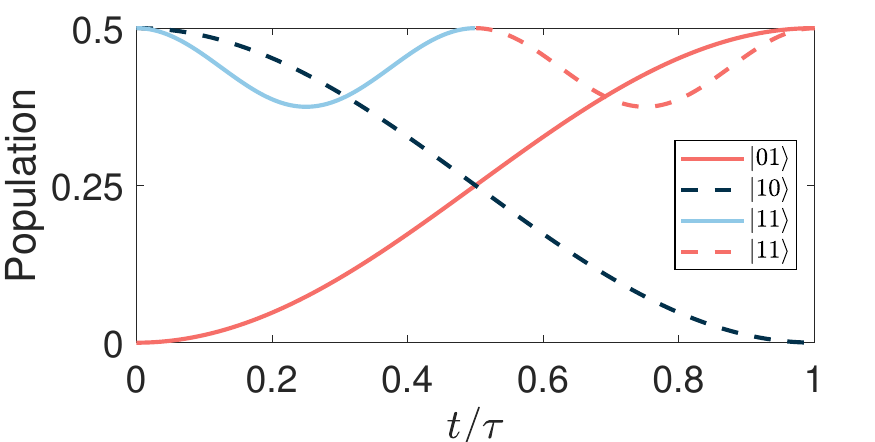}
 	\caption{Numerical simulation of fSim gate population corresponding to NNGQC+NHQC scheme. The blue solid line and red dashed line represent the population of $|11\rangle$ for the first and the second cycle and the red solid line and black dashed line stand for the population of $|01\rangle$ and $|10\rangle$ (the decoherence caused by environment is not considered in this simulation).}
 	\label{Fig3}
 \end{figure}

In order to obtain both evolutions simultaneously, it is only necessary to satisfy the evolution time $ n_1\times \tau_1= n_2\times \tau_2$ with $n_1,\  n_2\in N_+ $ ($N_+$ represents the set of positive integer natural numbers).  To minimize the impact of decoherence by minimizing the evolution time as much as possible, we have selected $n_1=1,\ n_2=2$, that is to say, $\alpha_A=\alpha_B=4\sqrt{3} \mathcal{G}$, i.e., $\tau=\tau_1=2\tau_2$. More specifically, $|11\rangle$ undergoes two cycles to obtain nonadiabatic holonomic phase 2$\phi$, and at the same time, $|01\rangle$ and $|10\rangle $ complete nonadiabatic nocyclic geometric iSWAP interaction. These evolutions can be described on the Bloch sphere as Fig.~\ref{Fig2}(b) and (c). It is worth noting that these two paths evolve simultaneously. And, by setting different values of detuning $\Delta$ and coupling strength $\mathcal{G}$, $|11\rangle$ can obtain different phase $\varphi$. To better represent two parallel evolutions, we plot the population of $|10\rangle,\ |01\rangle,\  |11\rangle$  of the initial state at $\frac{1}{\sqrt{2}}(|10\rangle+|11\rangle)$ in Fig.~\ref{Fig3}. The population colors of each state correspond to the colors in the Bloch spheres in Fig~\ref{Fig2}(b) and (c).

\section{Gate performance}\label{Sec4}
To fully analyze the feasibility of our scheme, we define the average fidelity as 
 \begin{eqnarray}
     \mathcal{F}=\frac{1}{4\pi^2}\int_0^{2\pi}\int_0^{2\pi} \langle \psi_f|\rho|\psi_f\rangle \rm d\theta_1 d \theta_2
 \end{eqnarray}
 with initial state $|\psi(0)\rangle=(\cos \theta_1 |0\rangle_1+\sin \theta_1 |1\rangle_1 )\otimes (\cos \theta_2 |0\rangle_2+\sin \theta_2 |1\rangle_2 )$, $|\psi_f\rangle$ is the ideal final state, and the density matrix of this system can be solved by Lindblad master equation
 \begin{equation}
         \dot{\rho}(t)=i \ [\rho(t),\hat{H}_s]+\sum_{j=1}^{2}[\frac{\kappa_-^j}{2}\mathcal{L}(\sigma_j)+\frac{\kappa_z^j}{2}\mathcal{L}(\chi_j)],
 \end{equation}
where $\mathcal{L}(\mathcal{A})=2\mathcal{A}\rho \mathcal{A}^\dagger-\mathcal{A}^\dagger\mathcal{A}\rho-\rho\mathcal{A}^\dagger\mathcal{A}$ is the Lindblad operator for operator $\mathcal{A}$, $\kappa_-^j$, $\kappa_z^j$ are the relaxation and dephasing rates of the $j$th Xmon, and $\sigma_j=|j-1\rangle \langle j|$ with $j\in \{1,\ 2\}$, $\chi_{j^\prime}=|j^\prime\rangle \langle j^\prime|$ for the $j^\prime$th level with $j^\prime\in \{ 0,\ 1,\ 2\}$ of the $j$th Xmon. From Fig~\ref{Fig4}, we can see the average fidelity of the NNGQC+NHQC method and conventional NGQC+NHQC~(detailed in the Appendix \ref{appB}) can achieve 0.9998 and 0.9997, respectively.

   \begin{figure}[htbp]
 	\centering \includegraphics[width=\linewidth]{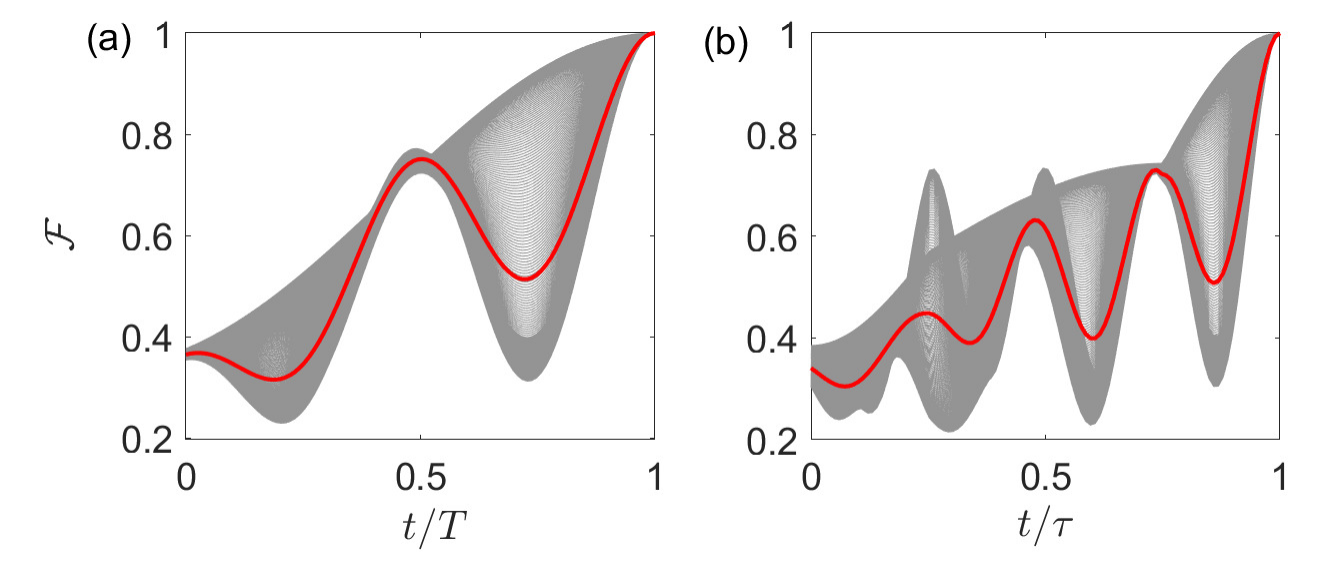}
 	\centering 
 	\caption{Numerical simulation of the average fidelity $\mathcal{F}$ of the fSim gate population~(red line) with NNGQC+NHQC~(a) and NGQC+NHQC~(b). The grey lines represent the fidelity of 500 different initial states, and the red line expresses the average fidelity of these 500 initial states. The parameter settings are as follows: $g=1$, $\alpha=4\sqrt{3}$, $\omega_ A=\omega_B=100$, and $\kappa_-^j=\kappa_z^j=10^{-4}$.}
 	\label{Fig4}
 \end{figure}

   \begin{figure}[htbp]
 	\centering \includegraphics[width=\linewidth]{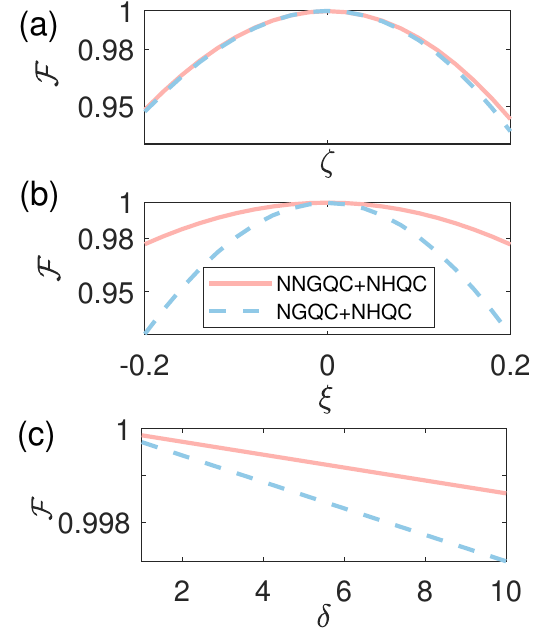}
 	\caption{The performance of fSim gate under parameters fluctuation. Numerical simulation of the average fidelity $\mathcal{F}$ of the fSim gate under the coupling strength error $\zeta$~(a), the frequency fluctuation of Xmon B~$\xi$~(b), and the decay~(dephasing) rate~$\delta$~(c).}
 	\label{Fig8}
 \end{figure}
Next, we will demonstrate the robustness of our scheme based on the methods NGQC+NHQC and NNGQC+NHQC against coupling strength errors and the frequency fluctuation of Xmon B.
We assume the coupling strength between Xmons A and B varies in the range of $\mathcal{G}\rightarrow(1+\zeta)\mathcal{G}$ with $\zeta\in[-0.2,0.2]$, and $\omega_B\rightarrow(1+\xi)\omega_B$ with $\xi\in[-0.2,0.2]$. Furthermore, we also simulate\sout{d} the average fidelity as a function of the relaxation rate and dephasing rate of the $j$th Xmon $\kappa_-^j=\kappa_z^j\rightarrow\delta \kappa_-^j=\delta \kappa_z^j$ with $\delta\in[1,10]$. From Fig.~\ref{Fig8}, we can see that even with a 20\% fluctuation in coupling strength $\mathcal{G}$, the fidelity of the two schemes can still be maintained above 0.93, and for the frequency fluctuation of Xmon B and the environment noise, the method of NNGQC+NHQC showcased greater superiority.

\section{Capacitive coupling via coupler}\label{Sec5}

The process of adjusting the qubit frequency may lead to an issue named ``frequency crowding'' and control crosstalk. Although the use of asymmetric transmons can help alleviate this problem, the impact of this issue still exists. A tunable coupler can help alleviate this problem~\cite{PhysRevApplied.10.054062}. Moreover, after adding a tunable coupler, the coupler strength of two logical qubits can be adjusted between exactly zero to over $100\ \rm MHz$ (absolute value), that is the coupling can be turned on and off by adjusting the coupler frequency $\omega_{c}$. Here, we will demonstrate our scheme can be performed in the circuit with a tunable coupler.
\begin{figure}[htbp]
	\centering \includegraphics[width=\linewidth]{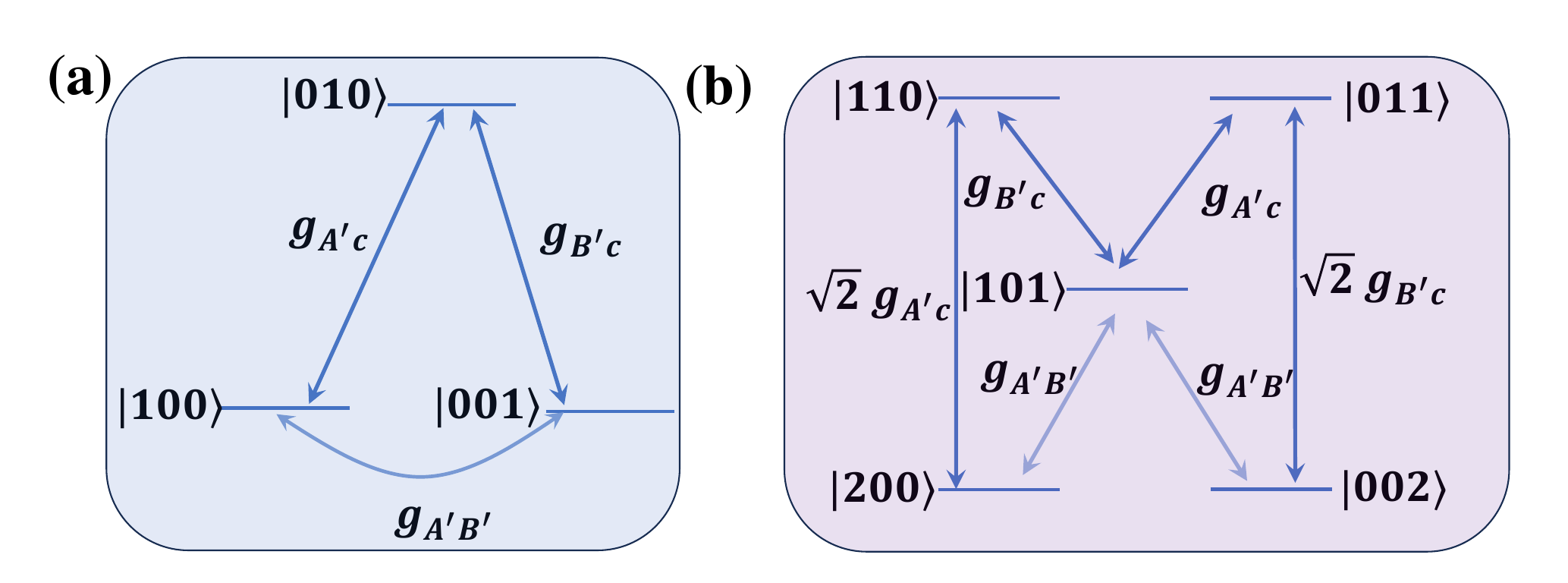}
	\caption{Energy level diagrams of the single-excitation manifold (a) and the double-excitation manifold (b) when performing the fSim gate. Blue double-headed arrows denote exchange interactions between the energy levels. The arrow color of $g_{A^\prime B^\prime}$ is lighter than the other to indicate the coupling of the nearest neighbor is stronger than the coupling of the next nearest neighbor.}
	\label{Fig5}
\end{figure}
Similar to the case of direct coupling, the Hamiltonian can be written in a form consistent with Eq.~(\ref{eq2})
\begin{eqnarray}
\label{eq4}
\hat{H}_{s^\prime}&=&\hat{H}_{A^\prime}+\hat{H}_{B^\prime}+\hat{H}_{c}+\hat{H}_{\rm int ^\prime},\notag\\
\hat{H}_{j^\prime}&=&\omega_{j^\prime}\hat{a}^\dagger_{j^\prime}\hat{a}_{j^\prime}-\frac{\alpha_{j^\prime}}{2}\hat{a}^\dagger_{j^\prime}\hat{a}^\dagger_{j^\prime}\hat{a}_{j^\prime}\hat{a}_{j^\prime}
\notag\\
\hat{H}_{\rm int^\prime}&=&\sum _{j < k} g_{jk}(\hat{a}^\dagger_j\hat{a}_k+\hat{a}_j\hat{a}^\dagger_k-\hat{a}^\dagger_j\hat{a}^\dagger_k-\hat{a}_j\hat{a}_k),\notag
\end{eqnarray}
where $j,k\in \{A^\prime,\ B^\prime,\ c\}$, and we set $A^\prime< B^\prime <c$, $\omega_{j^\prime }  $ is the transition frequency from ground state to the first excited state for the $j$th Xmon. Based on the actual situation, we consider that the system is at most doubly excited, and the energy level diagram is shown in Fig.~\ref{Fig5}.

To directly demonstrate the coupling between logical qubits, we adopt Schrieffer-Wolff transformation~\cite{BRAVYI20112793}
\begin{eqnarray}
\hat{U}=\sum_{m=A^\prime,B^\prime}&& [ \frac{g_{mc}}{\omega_m-\omega_c}(\hat{a}_m^\dagger\hat{a}_c-\hat{a}_c\hat{a}_m^\dagger)\notag\\&& -\frac{g_m}{\omega_m+\omega_c}(\hat{a}_m^\dagger\hat{a}_c^\dagger-\hat{a}_m\hat{a}_c)],
\end{eqnarray}
then, keep all terms to second order, we have
\begin{eqnarray}
    \hat{\widetilde{H}}&=&e^{\hat{U}}\hat{H}_{s^\prime} (e^{\hat{U}})^\dagger\notag\\
    &=&\sum_m \widetilde{\omega}_m^\prime\hat{a}_m^\dagger\hat{a}_m-\frac{\widetilde{\alpha}_m}{2}\hat{a}_m^\dagger\hat{a}_m^\dagger\hat{a}_m\hat{a}_m\notag\\
    && +\ \widetilde{g}(\hat{b}^\dagger_{A^\prime}\hat{b}_{B^\prime}+\hat{b}_{A^\prime}\hat{b}^\dagger_{B^\prime}), 
\end{eqnarray}
where
\begin{eqnarray}
 \widetilde{g} &\approx& \frac{g_{A^\prime c}g_{B^\prime c}}{2}\sum_m (\frac{1}{\omega_m-\omega_c}-\frac{1}{\omega_m+\omega_c})+g_{A^\prime B^\prime},\notag\\
\widetilde{\omega}_m&\approx&\omega_m + g_{mc}^2 (\frac{1}{\omega_m-\omega_c}-\frac{1}{\omega_m+\omega_c}),\notag\\
  \widetilde{\alpha}_m&\approx&\alpha_m.
\end{eqnarray}

\begin{figure}[htbp]
	\centering \includegraphics[width=\linewidth]{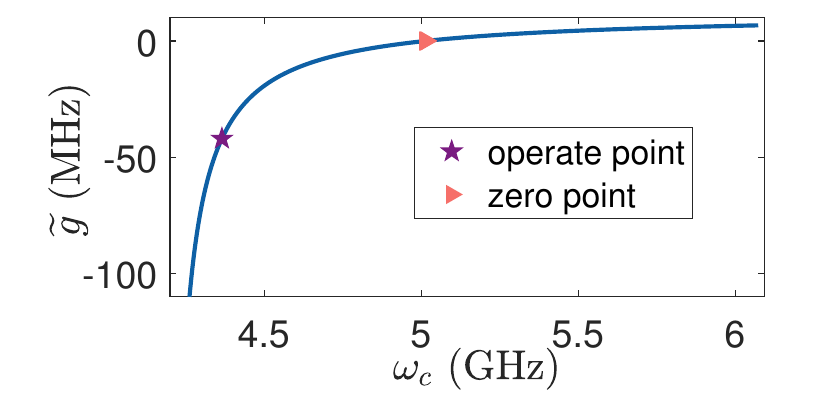}
	\caption{The effective coupling strength $\widetilde{g}$. At this point marked by a pentagram, the same coupling as direct coupling can be achieved. At this point marked by a triangle, the coupling between two logical qubits can be turned off.  }
	\label{Fig6}
\end{figure}
In this approximation, we assume that the coupler is always in the ground state.
We rotate the Hamiltonian into the interaction picture, and after using the RWA and second-order perturbation approximation, it can be observed that there is still only coupling between two logical qubits, that is to say, after adding a tunable coupler, the above scheme can still be achieved.

\section{Experimental feasibility}\label{Sec6}
\begin{figure}[htbp]
 	\centering \includegraphics[width=\linewidth]{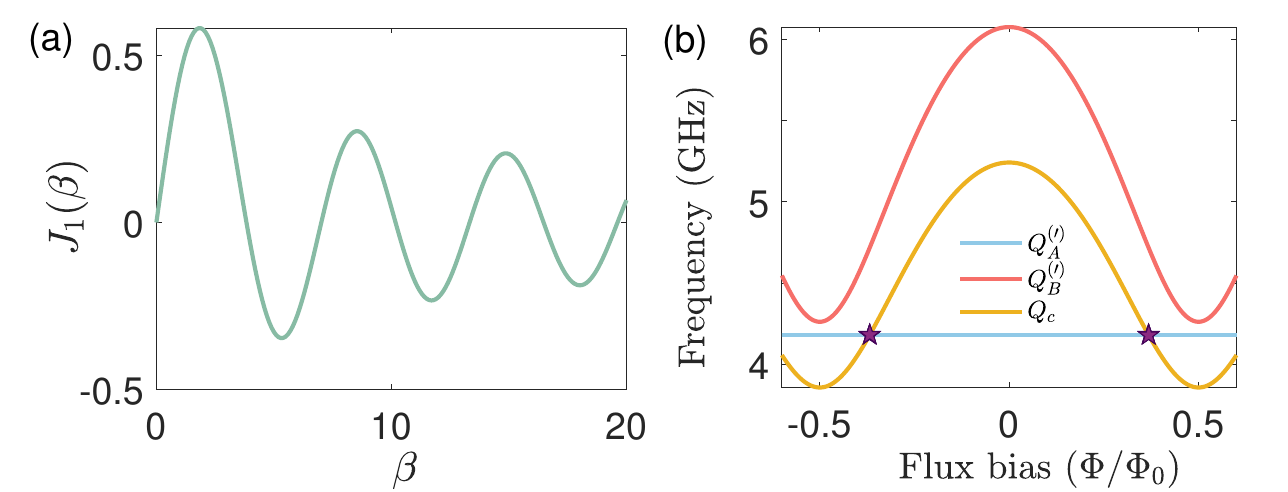}
 	\caption{(a) The Bessel function of the first kind varying $\beta$.(b) $|0\rangle\rightarrow|1\rangle$ transition frequencies of the logical qubit~(yellow and red) and the coupler~(red) varying with flux bias $\Phi/\Phi_0$. Relevant parameters are shown in Table.~\ref{tab1}.}
 	\label{Fig9}
 \end{figure}

In order to examine the feasibility of this scheme in the experiment, we now discuss the relevant parameters. Relevant parameters are shown in Table~\ref{tab1}, and based on these settings, the decay rate and the dephasing rate of each Xmon $\kappa_-^j=\kappa_z^j=2\pi\times 4.18 \ \rm kHz$. Based on these settings, the effective coupling strength between Xmon\textcolor{blue}{s} A and B is $\mathcal{G}=2\pi\times 41.8 \ \rm MHz$. Considering the Bessel functions of the first kind, $\nu= 0.369$, $\epsilon=0.692$, from Fig.~\ref{Fig9}(a), it can be seen that the coupling strength between Xmon\textcolor{blue}{s} A and B is $g=71.92\ \rm MHz$. Furthermore, as shown in Fig.~\ref{Fig9}(b), when the magnetic flux bias is adjusted to $\pm0.3153$, the frequency condition can be satisfied. The device parameters are demonstrated in Table \ref{tab1}. 
For the method of NGQC+NHQC and NNGQC+NHQC, the gate time is $23.92\ \rm ns$ and $11.96\ \rm ns$, respectively. 
These parameters are reasonable.

\begin{table}[htbp]
	\centering
	\caption{Device parameters.}
	\label{tab1}  
	\begin{tabular}{cccccc}
		\hline\hline\noalign{\smallskip}	
                                &	 & $Q_A$ &   & $Q_c$ & $Q_c$\\
		\noalign{\smallskip}\hline\noalign{\smallskip}
  	$E_c/2\pi$~($\rm GHz$)  &   &0.3 &  &0.12 &0.3  \\
        $E_j/2\pi$~($\rm GHz$)  &   &8.3627 &  &30 &10  \\
                                 &   &      &  &10 &2.8  \\
		$\omega/2\pi$~($\rm GHz$) &   &4.18 &  &$[4.262,~6.076]$&$[3.857,~5.241]$   \\
				$\alpha/2\pi $~($\rm MHz$)	  & & 300&  &120 &300 \\
		\noalign{\smallskip} \hline\hline
	\end{tabular}
\end{table}

\section{conclusion }\label{Sec7}
In conclusion, we have proposed a scheme to implement a geometric fSim gate with parallel paths in one step in a superconducting circuit. Our scheme can protect the fSim gate from the control error due to the intrinsic robustness of the geometric phase and the decoherence caused by the environment since the parallel paths shorten the evolution time.
Moreover, implementing the composite two-qubit gates in one step can greatly reduce the line depth in quantum simulation. In addition, our scheme does not require complex controls, making it very easy to implement in experiments, and can be achieved in various circuit structures. Therefore, our scheme may provide a significant reference and pave an alternative path for implementing low-depth fSim-gate-based quantum simulation in superconducting circuits.

 \section{Acknowledgement}
This work was supported by the National Natural Science Foundation of China under Grants (No. 12274376, No. U21A20434, No. 12074346), and a major science and technology project of Henan Province under Grant No. 221100210400, and the Natural Science Foundation of Henan Province under Grant No.  232300421075 and 212300410085, and Cross-disciplinary Innovative Research Group Project of Henan Province under Grant No. 232300421004.

 \appendix
 \section{Adjustable range of qubit frequency}\label{a1}

Here, we demonstrate how to adjust the qubit frequency by adjusting the magnetic flux passing through the SQUID.

Taking $Q_B$ as an example, the energy of the tunable Josephson energy can be described as 
\begin{widetext}
\begin{align}
E_{JB}&=E_{JBL}\ \cos{\phi_{BL}}+E_{JBR}\cos{\phi_{BR}}\notag\\
&= E_{JBL} \cos({\frac{\phi_{BL}+\phi_{BR}}{2}+\frac{\phi_{BL}-\phi_{BR}}{2}})+E_{JBR} \ \cos ({\frac{\phi_{BL}+\phi_{BR}}{2}-\frac{\phi_{BL}-\phi_{BR}}{2}})\notag\\
&=E_{JBL}(\cos\frac{\phi_{BL}+\phi_{BR}}{2}\cos\frac{\phi_{BL}-\phi_{BR}}{2}-\sin\frac{\phi_{BL}+\phi_{BR}}{2}\sin\frac{\phi_{BL}-\phi_{BR}}{2})\notag\\
& +E_{JBR}(\cos\frac{\phi_{BL}+\phi_{BR}}{2}\cos\frac{\phi_{BL}-\phi_{BR}}{2}+\sin\frac{\phi_{BL}+\phi_{BR}}{2}\sin\frac{\phi_{BL}-\phi_{BR}}{2})\notag\\
&=(E_{JBL}+E_{JBR})[\cos \frac{\phi_{BL}-\phi_{BR}}{2}\cos\frac{\phi_{BL}+\phi_{BR}}{2}]+(E_{JBR}-E_{JBL})[\sin \frac{\phi_{BL}-\phi_{BR}}{2}\sin\frac{\phi_{BL}+\phi_{BR}}{2}]\notag\\
&=E_{JB\Sigma}[\cos \frac{\pi \Phi}{\Phi_0}\cos \phi+d \sin \frac{\pi\Phi}{\Phi_0}\sin\phi]\notag\\
&= E_{JB\Sigma}\cos\frac{\pi\Phi}{\Phi_0}\sqrt{1+d^2\tan^2\frac{\pi\Phi}{\Phi_0}}\cos(\phi-\phi_0),
\end{align}
\end{widetext}
where $E_{JB\Sigma}=E_{JBL}+E_{JBR}$, $(\phi_{BL}+\phi_{BR})/2=\phi$, $\Phi_0=h/2e$ is the superconducting flux quantum, $d=({E_{JBL}-E_{JBR}})/({E_{JBL}+E_{JBR}})$, $\tan \phi_0=d \ \tan (\pi\Phi/\Phi_0)$, $\Phi$ is the magnetic flux passing through the SQUID, and $\phi_{JBL}-\phi_{JBR}=2\pi n +2\pi \Phi/\Phi_0$, $n\in N_+$.

 Therefore, the frequency of $Q_B$ is $\omega_B=\sqrt{8E_{JB}E_{CB}}-E_{CB}$.
 To better search for the physical quantities required for the experiment, we plot the frequency that varies with magnetic flux bias $\Phi/\Phi_0$.

\section{NGQC implementation of iSWAP part}\label{appB}
In this appendix, we present the details of implementing the iSWAP part with conventional NGQC.
We take the two-dimensional orthogonal eigenstates 
\begin{eqnarray}
    |\Psi_+(t)\rangle&=&\cos\frac{\theta(t)}{2}|01\rangle+\sin\frac{\theta(t)}{2}e^{i\chi(t)}|10\rangle\notag\\
     |\Psi_-(t)\rangle&=&\sin\frac{\theta(t)}{2}e^{i\chi(t)}|01\rangle-\cos\frac{\theta(t)}{2}|10\rangle
\end{eqnarray}
as our evolution states. To ensure the geometric of this evolution, the cyclic condition and the parallel-transport condition, i.e. 
\begin{eqnarray}
 (\rm i)\   && |\Psi_+(T)\rangle=e^{-i\gamma}|\Psi_+\rangle (0),\notag\\
 (\rm ii) \  && |\Psi_-(T)\rangle=e^{i\gamma}|\Psi_-\rangle (0),\notag\\
 (\rm iii)\   &&  \langle\Psi_\pm(t)|\hat{H}_r(t)|\Psi_\pm(t)\rangle=0,
\end{eqnarray}
that is, $|\Psi_\pm\rangle$ can obtain geometric phases $\pm \gamma$ without any dynamic phase at time $T$.

The evolution operator can be denoted as
\begin{eqnarray}
U_r(T)&=&e^{i\gamma}|\Psi_+(0)\rangle\langle\Psi_+(0)|+e^{-i\gamma}|\Psi_-(0)\rangle\langle\Psi_+(0)|\notag\\
    &=&\left(\begin{array}{cc}
      \cos\gamma-i\cos\theta\sin\gamma   & -ie^{i\chi}\sin\gamma\sin\theta \\
      -ie^{i\chi}\sin\gamma\sin\theta&     \cos\gamma+i\cos\theta\sin\gamma 
    \end{array}\right),\notag\\
\end{eqnarray}
where $\theta\equiv \theta(0)$, and $\chi\equiv \chi(0)$. When $\theta=\pi/2$, $\chi(0)=0$, and $\gamma=\pi/2$, an iSWAP evolution between $|01\rangle$ and $|10\rangle$ can be obtained, the coupling strength $\mathcal{G}$ and $\phi$ satisfying

    \begin{align}
\left \{
\begin{array}{ll}
\int_0^{T_1} \mathcal{G} \rm dt=\pi/4, \ \ \ \ \chi(t)=\pi/2,  & t\in [0, T_1],\\[2mm]
 \int_{T_1}^{T_2} \mathcal{G}\rm dt=\pi/2, \ \ \ \ \chi(t)=\pi ,     & t\in (T_1, T_2],\\[2mm]
\int_{T_2}^{T} \mathcal{G}\rm dt=\pi/4, \ \ \  \chi(t)=\pi/2  ,     & t\in (T_2, T].
\end{array}
\right.
\label{eq26}
\end{align}

\bibliography{manuscript}

\end{document}